\documentclass[10pt,twocolumn,letterpaper]{article}

\usepackage[pagenumbers]{cvpr} %

\usepackage{dsfont}
\usepackage{multirow}
\usepackage{diagbox}
\usepackage{graphicx}
\usepackage[normalem]{ulem}
\usepackage[dvipsnames]{xcolor}

\usepackage{soul}
\DeclareMathOperator{\unit}{unit}

\DeclareMathOperator{\Proj}{Proj}

\DeclareMathOperator{\GridSample}{GridSample}
\DeclareMathOperator{\vecgather}{gather}
\DeclareMathOperator{\argsort}{argsort}
\DeclareMathOperator{\LBS}{lbs}
\DeclareMathOperator{\diag}{diag}

\definecolor{cvprblue}{rgb}{0.21,0.49,0.74}
\usepackage[pagebackref,breaklinks,colorlinks,citecolor=cvprblue]{hyperref}

\def\paperID{24651246}
\def\confName{NoConf\xspace}
\def\confYear{1001\xspace}

\title{GIGA: Generalizable Sparse Image-driven Gaussian Humans}

\author{Anton Zubekhin\textsuperscript{1,2} \quad
Heming Zhu\textsuperscript{1} \quad
Paulo Gotardo\textsuperscript{3} \quad
Thabo Beeler\textsuperscript{3}\
\and
Marc Habermann\textsuperscript{1,2} \quad
Christian Theobalt\textsuperscript{1,2}\
\and
\small{\textsuperscript{1}
    Max Planck Institute for Informatics, Saarland Informatics Campus \quad}
\small{\textsuperscript{2}
    VIA Research Center \quad}
\small{\textsuperscript{3}
    Google}\
    \and
}

\begin{document}
\maketitle
\begin{abstract}
Driving a high-quality and photorealistic full-body virtual human from a few RGB cameras is a challenging problem that has become increasingly relevant with emerging virtual reality technologies.
A promising solution to democratize such technology would be a generalizable method that takes sparse multi-view images of any person and then generates photoreal free-view renderings of them. 
However, the state-of-the-art approaches are not scalable to very large datasets and, thus, lack diversity and photorealism.
To address this problem, we propose GIGA, a novel, generalizable full-body model for rendering photoreal humans in free viewpoint, driven by a single-view or sparse multi-view video.
Notably, GIGA can scale training to a few thousand subjects while maintaining high photorealism and synthesizing dynamic appearance.
At the core, we introduce a MultiHeadUNet architecture, which takes an approximate RGB texture accumulated from a single or multiple sparse views and predicts 3D Gaussian primitives represented as 2D texels on top of a human body mesh.
At test time, our method performs novel view synthesis of a virtual 3D Gaussian-based human from 1 to 4 input views and a tracked body template for unseen identities.
Our method excels over prior works by a significant margin in terms of identity generalization capability and photorealism.
\end{abstract}
    
\section{Introduction} 
\label{sec:intro}
Driving your own full-body and photoreal virtual double from a scarce and affordable single or sparse-view (up to 4) camera setup has the potential to revolutionize communication, gaming, and remote collaboration.
However, there are to date modeling challenges that remain unsolved: 
1) Achieving photorealism and fidelity despite sensor scarcity and limited input data.
2) Generalization ability to novel identities unseen during training.
In this work, we attempt to jointly solve them by leveraging recent large-scale data capture efforts. 
Notably, this requires a generalizable method to synthesize digital humans at test time in a simple feed-forward manner, which is the subject of this work.
\begin{figure}[t!]
    \centering
    \includegraphics[width=\linewidth]{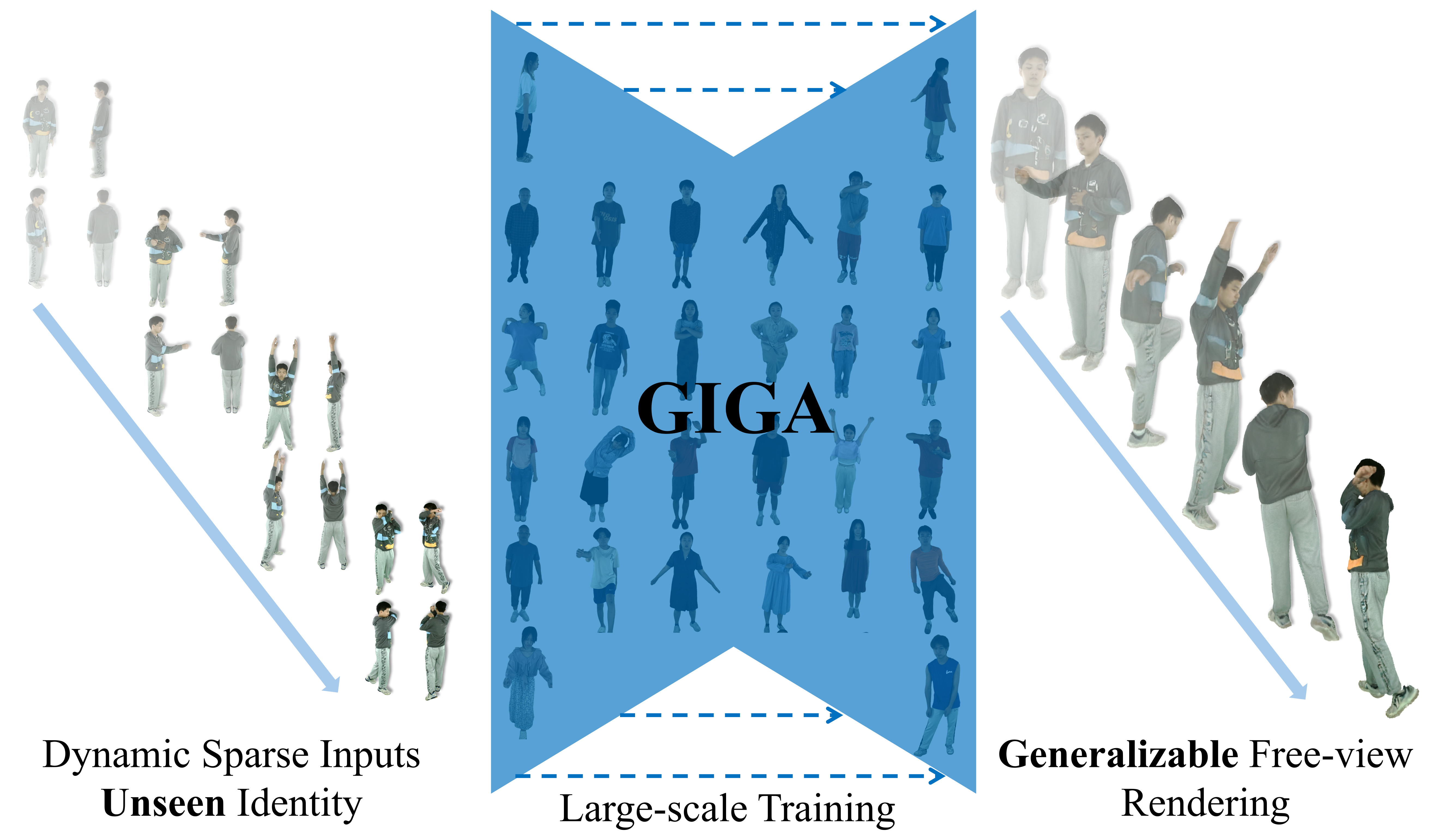}
    \caption{
    GIGA is trained efficiently on a large-scale human dataset. Given sparse views of an \textit{unseen} identity, and respective skeletal poses, GIGA generates photorealistic dynamic renderings in free viewpoint. 
    }
    \label{fig:teaser}
\end{figure}

\par
Recent related works have most focused on \textit{person-specific} avatars, i.e., a learned representation that is trained per subject on dense dome-like camera setups, which can be later articulated with arbitrary skeletal motion.
These representations may involve meshes \cite{habermann_2021_tog, bagautdinov_2021_tog, xiang_2022_tog}, neural radiance fields \cite{peng_2021_iccv, icsik_2023_tog, wang_2022_eccv,zhu2024trihuman}, points \cite{su_2023_iccv}, or volumetric primitives \cite{remelli_2022_siggraph} such as 3D Gaussians \cite{li2024animatable, pang_2024_cvpr, moreau_2024_cvpr}. 
Democratizing such technology is difficult since, for high-quality results, a dense camera dome is required \textit{before} the character can be driven at inference time.
Some methods aim to utilize simpler capture setups, e.g., monocular images \cite{xiu_2022_cvpr, xiu_2023_cvpr} or videos \cite{shaofeiwang_2024_cvpr, hushoukang_2024_cvpr}.
Nevertheless, due to the much scarcer input, their visual quality often falls short compared to multi-view methods. 
Single-image to 3D reconstruction methods~\cite{saito_2019_cvpr, saito_2020_cvpr, xiu_2022_cvpr, xiu_2023_cvpr, huang_2024_3dv, zhuang2025idol, dong2025moga} also primarily focus on adequate geometry reconstruction and produce a 3D asset, drivable with skeletal motion only, which typically lacks the skeletal motion-dependent geometry and appearance changes.
\par 
Designing a high-quality approach for lightweight human reconstruction and rendering, thus, remains an open problem:
the model must accurately represent diverse human appearances, body types, and clothing configurations, and it must correctly derive pose-dependent appearance changes from the scarce input signal.
While previous works \cite{kwon_2021_neurips, pan_2023_iccv, gao_2022_tvcg, kwon_2024_eccv} have shown promising first steps towards identity generalization, those methods were trained and evaluated on small-scale datasets \cite{peng_2021_cvpr} with few training subjects.
Moreover, the quality and efficiency of these works are constrained by the limitations of implicit neural representations and non-scalable network architectures.
More recently, large-scale datasets~\cite{cheng_2023_iccv, xiong_2024_cvpr} helped to pave the way for truly generalizable human modeling methods.
As was shown by~\citet{guo2025vid2avatar}, learning from a large data collection enables construction of a rich human avatar prior, albeit requiring preparation of personalized body templates beforehand.
Nonetheless, training and evaluating such models at scale demands particular consideration: 
First, a generalizable model has to learn meaningful feature representations to scale, both to the training set and outside of it. 
Second, the model architecture must be computationally and memory-efficient to capture fine detail and enable large-scale training.
Third, the underlying 3D representation has to yield high-quality, efficient rendering.
\par
To address these challenges, we propose GIGA, a feed-forward method to synthesize virtual 3D-Gaussian-based humans from single or sparse input views and a tracked body template at inference for free-view rendering.
Notably, for unseen subjects, GIGA \textit{requires no personalized training on dense dome data} and no fitting of personalized mesh templates.
At the core, we project sparse-view image features into the UV space of the SMPL-X model~\cite{pavlakos_2019_cvpr} while the virtual human predicted by GIGA is represented as a set of texel-aligned 3D Gaussians~\cite{kwon_2024_eccv,jiang_2024_arxiv, zhuang2025idol, yang2025sigman}.
We propose a MultiHeadUNet, a UNet with multiple encoding and decoding heads, which takes the texture containing the projected image information as well as shape and motion codes and regresses per-texel Gaussian appearance and geometry parameters.
In detail, we employ cross attention to inject motion information into the model, and skip-connections to propagate the learning signal at different spatial scales within the network.
Our architecture design choices ensure reliable learning of intrinsic feature statistics from the training data while maintaining person-specific information contained in the model inputs.
The predicted 3D Gaussians can be posed into 3D space using the respective SMPL-X body pose and further rendered onto novel views.
In summary, our contributions are:
\begin{itemize}
    \item We introduce GIGA, a novel generalizable method for dynamic novel view synthesis of virtual humans operating in a single or sparse view setup in a feed-forward regime at test time.
    \item We design
    a MultiHeadUNet with motion conditioning to synthesize high-quality texel-aligned 3D Gaussians with dynamic appearance changes.
    \item We train GIGA on a large-scale collection of multi-view videos, after which it generalizes to unseen identities within and across datasets, producing dynamic novel view renders even with as few as a single input video.
\end{itemize}
We demonstrate the generalization capabilities of GIGA through a comprehensive evaluation on large-scale multi-view~\cite{xiong_2024_cvpr, cheng_2023_iccv, tao_2021_cvpr} and monocular videos~\cite{jiang2022neuman} datasets.
Experimental results show that GIGA significantly outperforms prior works in visual quality as well as identity and pose generalization. 

\section{Related Work}
\label{sec:related}
\textbf{Person-specific Human Capture and Rendering.}
Neural implicit approaches for novel view synthesis \cite{mildenhall_2020_eccv, wang_2021_neurips, muller_2022_tog} were primarily designed for per-scene optimization.
This paradigm naturally extended to human modeling through various implicit representations: radiance fields \cite{peng_2021_cvpr, su_2021_neurips, icsik_2023_tog}, signed distance functions \cite{wang_2022_eccv, wang_2023_cvpr,zhu2024trihuman}, and occupancy fields \cite{saito_2019_cvpr, saito_2020_cvpr, mihajlovic_2021_cvpr}.
These methods typically rely on parametric human body models \cite{loper_2015_sigasia, pavlakos_2019_cvpr} as geometric proxies to initialize neural implicit representations.
The inherent limitations of generic parametric templates in capturing person-specific geometry led to methods utilizing personalized mesh templates \cite{liu_2020_tvcg, habermann_2021_tog, remelli_2022_siggraph}, which demonstrate superior reconstruction quality, mimicking pose-dependent appearance changes.
Drivable Volumetric Avatars (DVA) \cite{remelli_2022_siggraph} target telepresence applications by representing human avatars as mixtures of volumetric primitives \cite{lobardi_2021_tog}, regressed from texel-aligned image and pose features extracted from 3 input views.
Holoported Characters \cite{shetty_2024_cvpr} combines a personalized mesh template with dynamic feature textures predicted from partial texture and normal maps, operating with 4 input views, which yields 4K rendering resolution.
3D Gaussian Splatting \cite{kerbl_2023_tog} marked a significant advancement in both rendering quality and computational efficiency.
3D Gaussians have been successfully adapted for free-viewpoint human avatar rendering in multi-view \cite{pang_2024_cvpr, moreau_2024_cvpr, jiang_2024_arxiv, zielonka_2023_arxiv} and monocular \cite{hu_2024_CVPR, hushoukang_2024_cvpr, qian_2024_cvpr} setups, mostly constrained to person-specific optimization without cross-identity training.
\begin{figure*}[tp]
    \centering
    \includegraphics[width=\linewidth]{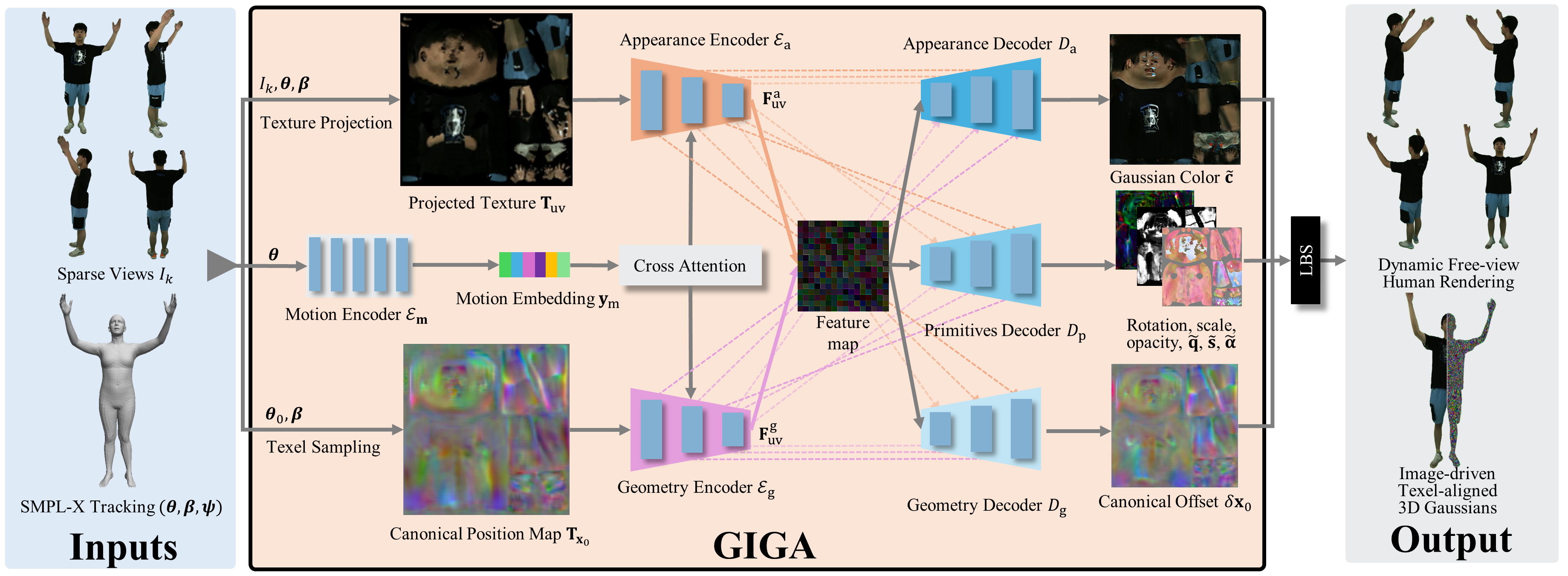}
    \caption{
    \textbf{Method Overview.} GIGA generates dynamic textures of 3D Gaussians for photoreal avatars from sparse input views $\mathcal{I}_k$ and a tracked body template $(\boldsymbol{\theta}, \boldsymbol{\beta})$. An initial RGB texture $\mathbf{T}_{\mathrm{uv}}$ is gathered from the input images and passed to the appearance encoder $\mathcal{E}_{\mathrm{a}}$ 
    to extract appearance features $\mathbf{F}^{\mathrm{a}}_{\mathrm{uv}}$. A canonical position map $\mathbf{T}_{\mathbf{x}_0}$ is processed by the geometry encoder $\mathcal{E}_{\mathrm{g}}$ into geometry features $\mathbf{F}^{\mathrm{g}}_{\mathrm{uv}}$. Both intermediate features are motion-dependent via cross-attention conditioning on the observed pose $\boldsymbol{\theta}$. 
    Gaussian texel maps are regressed by separate decoders, each for an individual group of parameters. 
    Skip-connections (dashed lines) propagate intermediate features from encoders to decoders.  
    Output of GIGA is articulated with linear blend skinning. 
    }
    \label{fig:method_pipeline}
\end{figure*}

\par \noindent \textbf{Generalizable Human Capture and Rendering.} \label{rel:generalizable_methods}
Recent research has addressed the challenge of synthesizing free-viewpoint videos of human performances from sparse multi-view captures with generalization across subjects~\cite{hu2024animate,zhu2024champ,huang2024magicfight}.
Neural Human Performer (NHP) \cite{kwon_2021_neurips} integrates pixel-aligned visual features with skeletal pose information extracted from multi-view video sequences through cross-attention.
TransHuman \cite{pan_2023_iccv} addresses performance degradation of NHP in occluded regions by learning to synthesize a NeRF-based representation in the canonical pose space with a transformer architecture that processes individual body parts as tokens.
Neural Novel Actor (NNA) \cite{gao_2022_tvcg} disentangles appearance and geometry by processing spatial point features and SMPL surface features through a graph CNN, enabling separate prediction of person-specific appearance and pose-dependent deformation, as proposed by \citet{liu_2021_sigasia}.
\par
More recently, GPS-Gaussian~\cite{zheng2024gps} achieved real-time novel view synthesis of human performances by predicting per-pixel 3D Gaussians from 2 source views, which are unprojected to 3D space and aggregated for final rendering.
Notably, GPS-Gaussian allows for view interpolation when source views have at most a 60-degree angular difference, thus requiring at least 6 input views in total.
Generalizable Human Gaussians (GHG) \cite{kwon_2024_eccv} introduces an alternative approach for reconstructing static human models from sparse input views.
By leveraging a large-scale dataset of textured human 3D scans \cite{tao_2021_cvpr}, GHG demonstrates generalization to unseen subjects by representing virtual humans as multiple scaffolds of 3D Gaussians in the UV space.
The most recent Vid2Avatar-pro~\cite{guo2025vid2avatar} achieved animatable avatar generation from a monocular video by training on a large corpus of multi-view data.
However, Vid2Avatar-pro requires a high-quality personalized body template at test time, which can only be reconstructed in a few hours for each novel identity.
\par \noindent \textbf{Generative Methods.}
\label{rel:generative_methods}
Another line of work focuses on single-image to animatable avatar reconstruction through generative modeling.
The most common technique here is to rely on SMPL-X and its UV mapping to anchor 3D Gaussians.
IDOL~\cite{zhuang2025idol} employs a transformer-based encoder to combine image features with a learnable UV token to later decode 3D Gaussians with a convolutional network.
Following IDOL, SIGMAN~\cite{yang2025sigman} learns a variational autoencoder~\cite{kingma2013auto} to map sparse views to 3D Gaussians, which is followed up by a diffusion transformer~\cite{peebles2023scalable} to generate latents for novel avatars, conditioned on an image or a text prompt.
Both IDOL and SIGMAN combine real-world datasets with synthetic data, generated with custom pipelines.
MoGa~\cite{dong2025moga} jointly optimizes a diffusion model to learn per-avatar latents and an autodecoder to convert latents to 3D avatars.
At test time, MoGa optimizes a latent code for the input image and uses a multi-view diffusion model to generate additional pseudo ground truth views for this latent inversion process.

\section{Method}
\label{sec:method}
In this section, we first formalize the problem setting and relevant background knowledge (Sec.~\ref{method:background}). 
Then we introduce our generalizable human representation (Sec.~\ref{method:feature_extraction}), our training strategy (Sec.~\ref{method:objective}), and finally the implementation details (Sec.~\ref{sec:implementation}).
\subsection{Problem Setting and Background}
\label{method:background}
\par \noindent \textbf{Problem Setting.}
GIGA aims at mapping single/sparse image observations (1-4 views) of an \textit{unseen} human to a 3D Gaussian representation that can be rendered from a free viewpoint, 
faithfully preserving the appearance and clothing dynamics from these driving videos, even for unseen identities. 
Fig.~\ref{fig:method_pipeline} presents an overview of our method.
To render dynamic sequences, GIGA processes each frame independently.
Following the common settings~\cite{remelli_2022_siggraph, shetty_2024_cvpr, kwon_2024_eccv}, we assume that a motion tracking and a parametric body template are given to roughly describe the body shape and the pose of the human.
Importantly, our method requires {\it no personalized training}. 
\par
For training, GIGA assumes a collection of multi-view videos of several hundred subjects with per-frame subject segmentation masks.
Each subject is captured by at least $\hat{K}=16$ calibrated cameras, with $\pi_{\hat{k}}$ denoting the projection matrix for camera $\hat{k}$.
Each video frame is annotated with SMPL-X~\cite{pavlakos_2019_cvpr} parameters released as additional labels in most multi-view human performance capture datasets ~\cite{xiong_2024_cvpr, cheng_2023_iccv, tao_2021_cvpr}.
During training, we randomly sample 3 or 4 input views and use estimated SMPL-X parameters to construct the input for the encoder network.
Images from the remaining views serve as ground truth to supervise the outputs decoded by the network.
\par \noindent \textbf{3D Gaussian Splatting (3DGS).} 
3DGS \cite{kerbl_2023_tog} models a scene using 3D Gaussian primitives. Each primitive
$\mathcal{G} = \{\boldsymbol{\mu},\mathbf{\Sigma}, \alpha, \mathbf{c}\}$ is paramterized by its position $\boldsymbol{\mu} \in \mathbb{R}^3$, 
covariance matrix $\mathbf{\Sigma} \in \mathbb{R}^{3 \times 3}$, 
opacity $\alpha \in \mathbb{R}$, 
and RGB colors $\mathbf{c} \in \mathbb{R}^3$.
The covariance matrix $\mathbf{\Sigma}$ is factorized as $\mathbf{\Sigma} = \mathbf{R} \mathbf{S} \mathbf{S}^T \mathbf{R}^T$, 
where the rotation $\mathbf{R}$ matrix is obtained from the quaternion $\mathbf{q} \in \mathbb{R}^4$ and the diagonal scaling matrix $\mathbf{S}=\diag\left(\mathbf{s}\right)$, with per-axis scales $\mathbf{s} \in \mathbb{R}^3$.
The Gaussians $\mathcal{G}$ is then rendered from the target camera $\pi$ to an image $I$ and an accumulated density image $A$ using a Gaussian rasterizer 
\begin{equation}
    I, A = \mathcal{R}(\mathcal{G}, \pi).
    \label{eq:rasterizer}
\end{equation}
\par \noindent \textbf{SMPL-X Human Body Template.} 
SMPL-X~\cite{pavlakos_2019_cvpr} is a parametric human body model with articulated limbs, hands, and an expressive face. 
SMPL-X has $N=10475$ vertices, $J=54$ joints, and can be articulated using linear blend skinning~\cite{lewis_2023_sgp} to obtain a set of posed vertices $\mathcal{V} \in \mathbb{R}^{N \times 3}$.
Formally, SMPL-X is defined as a function $\mathcal{V} = \mathcal{M}(\boldsymbol{\theta}, \boldsymbol{\beta}, \boldsymbol{\psi})$, with $\boldsymbol{\theta} \in \mathbb{R}^{(J+1) \times 3}$ describing $J$ joint angles and a rigid root transformation, body shape $\boldsymbol{\beta} \in \mathbb{R}^{10}$, and face expression $\boldsymbol{\psi} \in \mathbb{R}^{10}$. 
As we tested GIGA without expression tracking, we fixed $\boldsymbol{\psi}$ to the neutral expression.
\subsection{Generalizable Human Representation}
\label{method:feature_extraction}
\begin{figure*}[tp]
    \centering
    \includegraphics[width=1\linewidth]{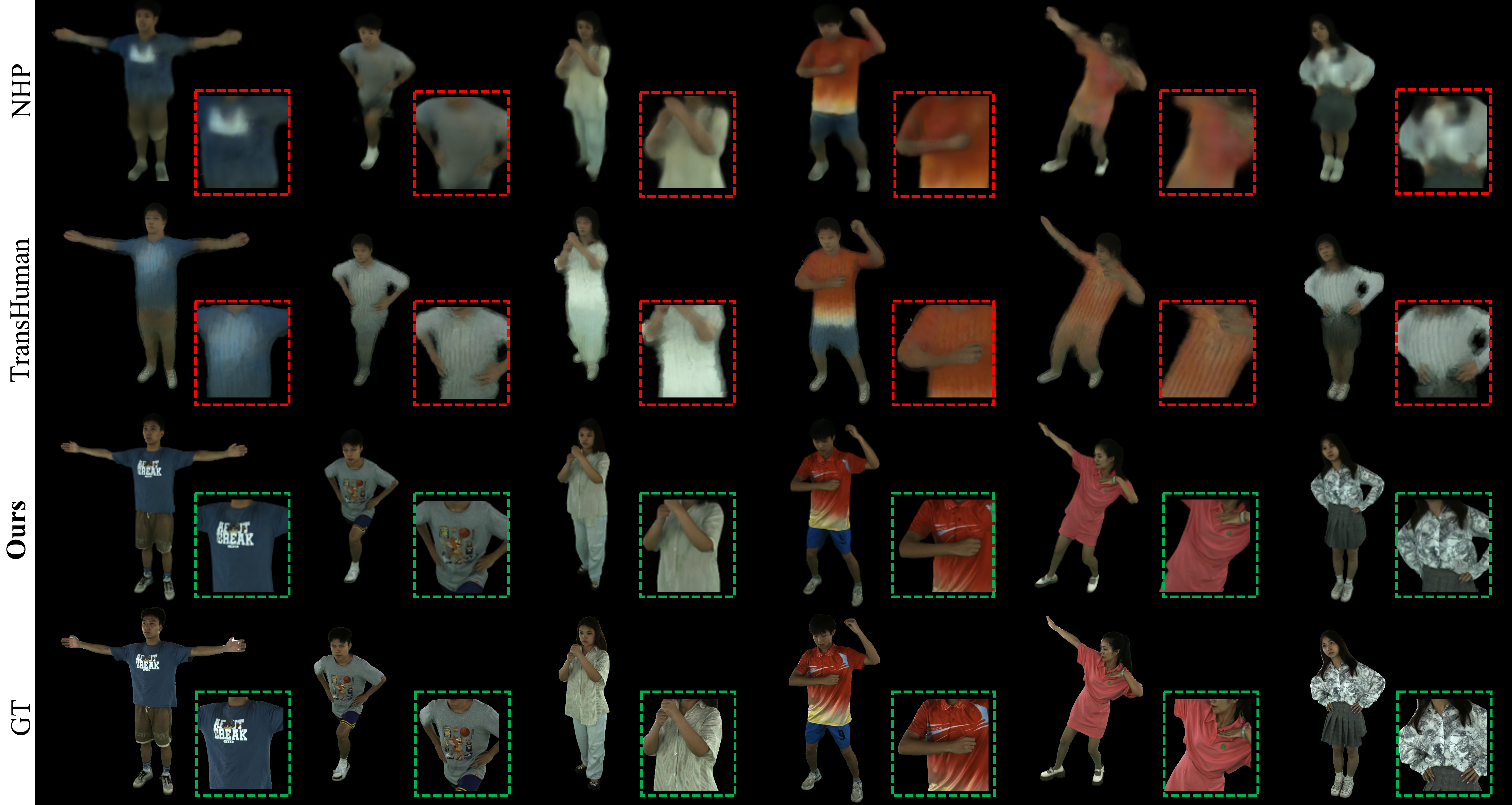}
    \caption{\textbf{Qualitative Comparison against Dynamic Methods.} We show results for identity generalization. GIGA (Ours) achieves significantly higher quality, while also being able to faithfully synthesize a virtual avatar after training on a large-scale dataset \cite{xiong_2024_cvpr}.}
    \label{fig:main_baseline}
\end{figure*}

Inspired by previous \textit{personalized}, animatable Gaussian avatar methods~\cite{pang_2024_cvpr, jiang_2024_arxiv, Wang_2025_arxiv_relcodeca},
we represent virtual humans as texel-aligned 3D Gaussian maps within the texture space $\mathcal{M}_{\mathrm{uv}}$ of the SMPL-X template mesh, displacing each 3D Gaussian in the canonical pose with a motion-dependent offset for better geometry modeling.
Texel-aligned 3D Gaussians are semantically similar across various characters and
2D textures can be efficiently processed by convolutional neural networks.
Thus, the predicted Gaussians from the same region of the texture space will be entangled thanks to the locality bias of convolutions. 
As such, GIGA can learn cross-subject information and predict the final texel-aligned 3D Gaussians with a single pass of a 2D convolutional neural network.
\par \noindent \textbf{Input Appearance Encoding.} 
We use information from the input views to capture identity-specific appearance with pose-dependent variations.
To transform this information from image to texture space, we first compute partial textures $\mathbf{T}_{\mathrm{uv}, k}$ for each view $\pi_{k}$ by projecting each pixel onto the 3D surface of SMPL-X. 
This 3D point can be mapped into texture space using the UV mapping of SMPL-X such that the final image pixel color is projected onto the respective 2D location in the texture map.
We then fuse those textures according to visibility, resulting in the final texture $\mathbf{T}_{\mathrm{uv}} \in \mathbb{R}^{T \times T \times 3}, T=512$ in the UV space $\mathcal{M}_{\mathrm{uv}}$.
\par
This texture is the input to our appearance encoder $\mathcal{E}_{\mathrm{a}}$:
\begin{equation}
    \mathbf{F}_{\mathrm{uv}}^{\mathrm{a}}, \mathbf{H}^{\mathrm{a}} = \mathcal{E}_{\mathrm{a}} \left(\mathbf{T}_{\mathrm{uv}}; \mathbf{y}_m \right) \,,
    \label{eq:method:appr}
\end{equation}
which extracts appearance features $\mathbf{F}_{\mathrm{uv}}^{a} \in \mathbb{R}^{T_f \times T_f \times d}$, that encode character-specific appearance and identity information.
The encoder $\mathcal{E}_\mathrm{a}$ consists of 2D convolutional downsampling residual blocks.
Feature maps $\mathbf{H}^{\mathrm{a}}$ from also taken from the downsampling levels for later usage in decoding.
$\mathcal{E}_\mathrm{a}$ is conditioned on the motion embedding $\mathbf{y}_{\mathrm{m}}$.  
Following~\citet{rombach2021highresolution}, this conditioning is implemented using cross-attention blocks as final layers of the encoder.
\par \noindent \textbf{Motion Embeddings.} 
Since the input texture $\mathbf{T}_{\mathrm{uv}}$ contains only a localized pose-dependent learning signal from the template surface.
We additionally use an MLP-based motion encoder $\mathcal{E}_\mathrm{m}$ to construct motion embeddings for SMPL-X poses $\boldsymbol{\theta}$, which adds global body motion awareness at the encoding stage:
\begin{equation}
    \mathbf{y}_{\mathrm{m}} = \mathcal{E}_{\mathrm{m}}\left(\boldsymbol{\theta}\right).
    \label{eq:motion_encoder}
\end{equation}
\par \noindent \textbf{Input Geometry Encoding.}
Even though appearance information is texel-aligned, it is insufficient to infer correct human shapes.
To address this, we employ a geometry encoder $\mathcal{E}_\mathrm{g}$ to extract approximate geometry information from the body template.
Starting with a T-posed SMPL-X mesh $\mathcal{V}(\boldsymbol{\theta}_0, \boldsymbol{\beta})$,
we compute a canonical position map $\mathbf{T}_{\mathbf{x}_0} \in \mathbb{R}^{T \times T \times 3}$ from canonical vertex coordinates $\mathbf{x}_0\left(\boldsymbol{\beta}\right) \in \mathbb{R}^{N \times 3}$ using the UV map $\mathcal{M}_{\mathrm{\mathrm{uv}}}$.
The geometry encoder $\mathcal{E}_{\mathrm{g}}$ has the same architecture as the appearance encoder $\mathcal{E}_\mathrm{a}$ and produces geometry features $\mathbf{F}_{\mathrm{uv}}^{\mathrm{g}} \in \mathbb{R}^{T_f \times T_f \times d}$ with a stack of feature maps $\mathbf{H}^{\mathrm{g}}$:
\begin{equation}
    \mathbf{F}_{\mathrm{uv}}^\mathrm{g}, \mathbf{H}^{\mathrm{g}} = \mathcal{E}_\mathrm{g} \left(\mathbf{T}_{\mathbf{x}_0}; \mathbf{y}_{\mathrm{m}}\right)
    \label{eq:method:geom}.
\end{equation}
To handle dynamically changing details in the final shape, the geometry encoder $\mathcal{E}_\mathrm{g}$ is also conditioned on the motion embedding $\mathbf{y}_{\mathrm{m}}$.
\par \noindent \textbf{Gaussian Primitives Regression.}
For the decoding stage, we design three separate decoders (Fig.~\ref{fig:method_pipeline}): 
$\mathcal{D}_\mathrm{a}$ for the appearance, 
$\mathcal{D}_\mathrm{p}$ for scales, quaternions and opacities of 3D Gaussians, 
and $\mathcal{D}_\mathrm{g}$ for per-Gaussian offsets.
All three share the same convolutional architecture and receive appearance and geometry features $\mathbf{F}_{\mathrm{uv}}^\mathrm{a}, \mathbf{F}_{\mathrm{uv}}^\mathrm{g}$ as inputs:
\begin{equation}
    \mathcal{G}'_{\mathrm{uv}} =
    \mathcal{D}_{\{ \mathrm{a} | \mathrm{p} | \mathrm{g}\}} \left([\mathbf{F}_{\mathrm{uv}}^\mathrm{a}, \mathbf{F}_{\mathrm{uv}}^\mathrm{g}]; [\mathbf{H}^\mathrm{a}, \mathbf{H}^\mathrm{g}]\right).
    \label{method:eq:decoders}
\end{equation}
The output map $\mathcal{G}'_{\mathrm{uv}}$ contains RGB color channels $\mathbf{c}$, quaternions $\mathbf{q}_0$, normalized scales $\mathbf{s'}$, opacities $\alpha$, and offsets $\delta \mathbf{x}_0$, which are defined in the canonical T-pose.
To utilize the representational power of the shared texel space and to propagate semantic information from encoders to decoders at different spatial scales, 
we use UNet-like~\cite{ronneberger_2015_unet} skip-connections, building a feature propagation bridge from each encoder to each decoder.
We stack intermediate feature maps $\mathbf{H}^\mathrm{a}$ and $\mathbf{H}^\mathrm{g}$ along the feature dimension and feed them to every decoder at each corresponding upsampling layer.
To convert a texel-aligned 3D Gaussian map to the observed pose space $\boldsymbol{\theta}$, we use linear blend skinning applied to both offsets $\delta \mathbf{x}_0$ and quaternions $\mathbf{q}_0$: $\delta \mathbf{x} = \LBS\left(\delta \mathbf{x}_0, \boldsymbol{\theta}\right); \mathbf{q} = \LBS \left(\mathbf{q}_0, \boldsymbol{\theta}\right)$.
We also multiply normalized Gaussian scales $\mathbf{s}'$ by a hyperparameter $\rho$: $\mathbf{s} = \rho \mathbf{s}'$, value $\rho=5e^{-3}$ was empirically found adequate to provide sufficient coverage of the template with Gaussians.
\par
The resulting 3D Gaussian map $\mathcal{G}_{\mathrm{uv}}$ is rendered for each camera view $\pi_{k}$ using differentiable Gaussian rasterizer~\cite{ye_2024_arxiv}, yielding an RGB and accumulated opacity image pair, $I_k, A_k = \mathcal{R}\left( \mathcal{G}_{\mathrm{uv}}, \pi_k \right)$, as in Eq.~\eqref{eq:rasterizer}.
\subsection{Training}
\label{method:objective}
GIGA generates texel-aligned 3D Gaussian texture maps at $512 \times 512$ resolution.
As some texels in these textures are not associated with any triangle, they are assigned zero opacity during rendering.
\par \noindent \textbf{Training Objective.}
GIGA is trained to minimize a combination of the reconstruction $\mathcal{L}_{\mathrm{rec}}$ and geometry-related regularization terms $\mathcal{L}_{\mathrm{reg}}$:
\begin{equation}
    \mathcal{L} = \mathcal{L}_{\mathrm{rec}} + \mathcal{L}_{\mathrm{reg}}\,,
    \label{loss:full}
\end{equation}
\par \noindent \textbf{Reconstruction Term.}
Following~\citet{kerbl_2023_tog}, we compute the mean absolute error $\mathcal{L}_{\mathrm{L1}}$ and the structural similarity measure $\mathcal{L}_{\mathrm{ssim}}$ \cite{wang_2004_ieee} between the rendered image $I_k$ and the ground-truth image $I_{\mathrm{gt}, k}$.
We additionally evaluate the VGG-based \cite{simonyan_2015_arxiv} perceptual loss \cite{zhang_2018_cvpr} $\mathcal{L}_{\mathrm{LPIPS}}$ on randomly sampled patches with centers within the ground truth segmentation mask $A_{\mathrm{gt}, k}$. 
To improve outlines of Gaussian primitives, we also compute $\mathcal{L}_{\mathrm{mask}}$, the mean squared error between the rendered opacity images $A_k$ and ground truth character segmentation masks $A_{\mathrm{gt}, k}$.
The final reconstruction term is averaged over all images in the batch and defined as:
\begin{equation}
    \mathcal{L}_{\mathrm{rec}} = \lambda_1 \mathcal{L}_{\mathrm{L1}} + \lambda_2 \mathcal{L}_{\mathrm{ssim}} + \lambda_3 \mathcal{L}_{\mathrm{LPIPS}} + \lambda_4 \mathcal{L}_\mathrm{mask} \,,
    \label{loss:recon}
\end{equation}
with $\lambda_1 = \lambda_2 = \lambda_3 = 0.5$ and $\lambda_4 = 0.1$ in all experiments.
\begin{figure}[tp]
    \centering
    \includegraphics[width=\linewidth]{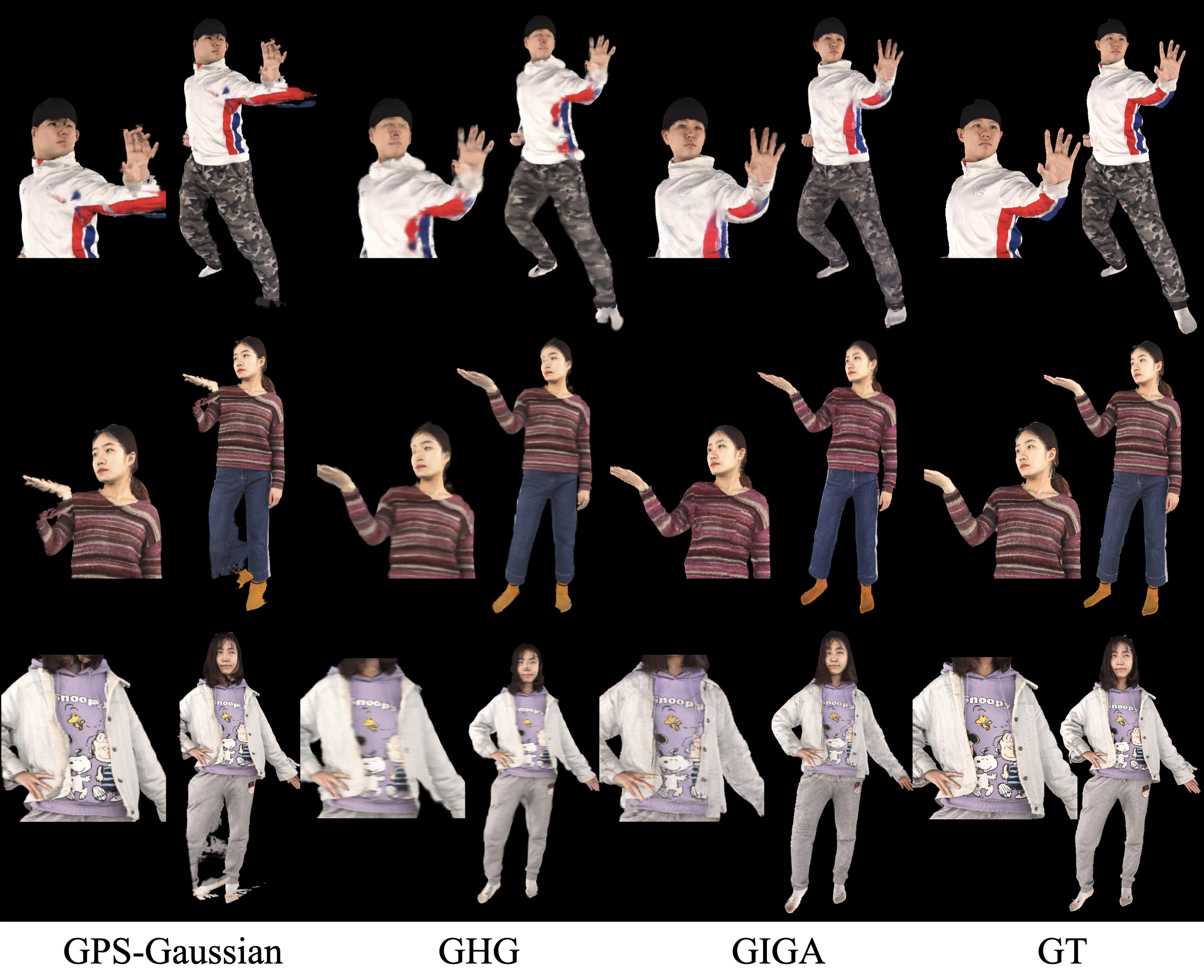}
    \vspace{-8pt}
    \caption{\textbf{Qualitative Comparison against 3D Gaussian-based methods.} GIGA achieves higher rendering quality than previous GHG~\cite{kwon_2024_eccv} and GPS-Gaussian~\cite{zheng2024gps} with fewer artifacts. Results reported on THuman2.1 dataset~\cite{tao_2021_cvpr}.}
    \vspace{-8pt}
\label{fig:ghg_baseline}
\end{figure}

\par \noindent \textbf{Regularization Term.}
Modeling various dynamically changing geometrical shapes with Gaussian offsets $\delta \mathbf{x}_0$ is particularly challenging during the early stages of optimization.
To this end, we introduce additional penalties to constrain some of the predicted Gaussian parameters:
\begin{equation}
\begin{aligned}
    &\ \mathcal{L}_{\mathrm{reg}} = \lambda_5 \mathcal{L}_{\delta \mathbf{x}_0} + \lambda_6 \left(\mathcal{L}_{\delta\mathbf{x}_0; \alpha} + \mathcal{L}_{\mathbf{s}'; \alpha}\right) + \lambda_7 \mathcal{L}_\alpha\ = \\
    &\ = \lambda_5 \left\lVert \delta \mathbf{x}_{0} \right\rVert_2 + \lambda_6 \left(\left\lVert \mathds{1}_{\left[\alpha < \epsilon\right]} \mathbf{x}_{0} \right\rVert_2 + \left\lVert \mathds{1}_{\left[ \alpha < \varepsilon \right]}\mathbf{s}'\right\rVert_2\right) + \\
    &\ + \lambda_7 \textrm{Beta}\left(\alpha\right)\,,
    \label{loss:regularizers}
\end{aligned}
\end{equation}
with $\lambda_5=0.15, \lambda_6=\lambda_7=0.1$. 
The offset penalty $\mathcal{L}_{\delta \mathbf{x}_0}$ prevents Gaussians from drifting far away from the mesh.
\par
Due to UV mapping distortions, some mesh regions may be oversampled with Gaussians. The poor initial model state will encourage some Gaussians to become transparent and inadequately large and move away from the mesh. 
To reclaim those Gaussians for actual modeling, we additionally penalize offsets of low-opacity primitives with $\mathcal{L}_{\delta \mathbf{x}_0; \alpha}$,
where the indicator function $\mathds{1}_{\left[\alpha < \epsilon\right]}$ triggers the extra penalty only when Gaussian opacity $\alpha$ is below a threshold $\epsilon$. Importantly, we disable gradient flow through the indicator function.
We similarly penalize scales $\mathbf{s}$ when opacity is below $\varepsilon$ with $\mathcal{L}_{\mathbf{s}'; \alpha}$.
Following~\citet{Lombardi_2019_sig}, we encourage all Gaussians to be either completely opaque or completely transparent by computing the negative log-likelihood of the beta distribution $\textrm{Beta}\left(0.5, 0.5\right)$ as $\mathcal{L}_\alpha$.
\subsection{Implementation Details} \label{sec:implementation}
We train GIGA for $250$k iterations on a single NVIDIA H100 GPU using 4 subjects per batch. 
All models are optimized with AdamW \cite{loshchilov_2017_iclr}, with the learning rate being linearly increased from $0$ to $1e^{-4}$ during the first $25$k training steps. We also set the weight decay parameter to $1e^{-4}$.
\par
At the warmup stage, we also apply linear annealing to the predicted offsets $\delta \mathbf{x}_0$ in addition to other penalties, Eq.\eqref{loss:regularizers}.
We empirically find it crucial, as it allows the model to focus on approximate appearance reconstruction before refining person-specific geometry, effectively preventing unstable offset predictions.
\begin{table}[t]
    \centering
    \begin{tabular}{l|l|c|c|c}
    \hline
    \cline{2-5}
    Dataset & Method & $\uparrow$ PSNR & $\uparrow$ SSIM & $\downarrow$ LPIPS \\
    \hline
    \multirow{3}{*}{{\footnotesize MVHumanNet}} & NHP & 17.72 & 0.5678 & 402.7 \\
    & TH & 17.46 & 0.5808 & 392.1 \\
    & \textbf{GIGA}  & \textbf{22.19} & \textbf{0.7526} & \textbf{70.2} \\
    \hline
    \multirow{3}{*}{{\footnotesize THuman2.1}} & GPS-G & 18.22 & 0.7768 & 207.9 \\
    & GHG & 18.86 & 0.7394 & 76.5 \\
    & \textbf{GIGA}  & \textbf{20.19} & \textbf{0.7944} & \textbf{58.7} \\
    \hline
    \end{tabular}
    \caption{\textbf{Quantitative Comparison.} We evaluate generalization to unseen identities of Neural Human Performer (NHP) and TransHuman (TH) on MVHumanNet; GPS-Gaussian (GPS-G) and Generalizable Human Gaussians (GHG) on THuman2.1.
    }
    \label{tab:baselines}
\end{table}

\section{Results}
\label{sec:results}
We first explain the datasets and metrics (Sec.~\ref{sec:dataset}).
Then, we show qualitative results (Sec.~\ref{sec:qualitative}) and comparisons (Sec.~\ref{sec:comparison}).
Lastly, we ablate our design choices (Sec.~\ref{sec:ablations}).
Supplementary materials contain additional qualitative examples.
\subsection{Dataset and Metrics} \label{sec:dataset}
\par \textbf{Datasets.} \label{sec:dataset}
To evaluate GIGA, we use the multi-view human performance capture datasets MVHumanNet~\cite{xiong_2024_cvpr} and DNA-Rendering~\cite{cheng_2023_iccv}, and also the 3D human scan dataset THuman2.1~\cite{tao_2021_cvpr}. 
Details on the data usage protocol for training can be found in the supplementary material.
\par \noindent\textbf{Metrics.}
\label{sec:metrics}
To evaluate rendering quality, we compute the following metrics: PSNR, SSIM~\cite{wang_2004_ieee}, and the perceptual LPIPS~\cite{zhang_2018_cvpr} using AlexNet~\cite{krizhevsky_2017_acm} features (scaled by 1000).
\subsection{Qualitative Results}
\label{sec:qualitative}
Fig.~\ref{fig:main_baseline},~\ref{fig:ghg_baseline}, and \ref{fig:cross_dataset} (and the supplementary video) present GIGA's novel view rendering results of unseen subjects performing unseen poses.
Notably, GIGA achieves photorealistic and view-consistent rendering and effectively captures fine details such as clothing wrinkles and intricate textures.
We highlight that for novel subjects, GIGA achieves a rendering quality comparable to that of the training subjects, demonstrating its generalization ability to novel identities.
Please see the in-the-wild examples from Neuman dataset in the supplementary materials.
\begin{table}[t]
\centering
    \centering
    \begin{tabular}{l|c|c}
    & \multicolumn{2}{c}{Training set $\to$ Testing set} \\
    \hline
    Metric & DNA $\to$ DNA & MVH $\to$ DNA \\
    \hline
    $\uparrow$ PSNR & 19.63 & 19.49 \\
    \hline
    $\uparrow$ SSIM & 0.7255 & 0.7138 \\
    \hline
    $\downarrow$ LPIPS & 106.4 & 114.9 \\
    \hline
    \end{tabular}
    \caption{\textbf{Generalization to Novel Data.} Training on MVHumanNet allows GIGA to generalize to DNA-Rendering with minor quantitaive performance drop in comparison with GIGA trained on DNA-Rendering.}
    \label{tab:cross_data_generalization}
\end{table}

\subsection{Comparison}
\label{sec:comparison}
\noindent\textbf{Competing Methods.}
\label{sec:competing}
We compare GIGA to other generalizable dynamic image-driven methods discussed in Sec.~\ref{rel:generalizable_methods}, i.e., NHP~\cite{kwon_2021_neurips}, TransHuman (TH)~\cite{pan_2023_iccv}, GPS-Gaussian~\cite{zheng2024gps}, and GHG~\cite{kwon_2024_eccv}.
We exclude concurrent human avatar reconstruction methods IDOL~\cite{zhuang2025idol}, SIGMAN~\cite{yang2025sigman}, and MoGa~\cite{dong2025moga} from comparison, since they focus on reconstructing animatable avatars from  single images.
\par 
Tab.~\ref{tab:baselines} provides quantitative comparisons of GIGA against state-of-the-art baselines for generalizable human rendering.
In all cases, we evaluate the generalization ability to unseen identities from held-out validation sequences.
Fig.~\ref{fig:main_baseline} provides an overview of qualitative results.
NHP relies on sparse 3D convolutions to process volumetric features in the observed pose space and suffers from missing input signals due to occlusions, ultimately failing to generalize to unseen identities.
Despite operating in a canonical template pose space and tokenizing the template for processing with a transformer-based network,
TransHuman also cannot learn meaningful priors from the large data collection.
GIGA, on the other hand, utilizes the power of the shared texel space to a maximum degree: all feature representations for digital humans are defined in the same texel space, and intermediate features enhance the quality of the final prediction through skip-connections.
MultiHeadUNet is also significantly more computationally efficient, which explains both qualitative and quantitative improvements over baselines.
\par
GPS-Gaussian heavily relies on depth estimation from two adjacent source views to unproject pixel-aligned 3D Gaussians for novel view rendering.
As such, it is limited by the quality of the depth estimation, which might fail if source views are too far from each other.
Artifacts of erroneous depth-based unprojection are visible in Fig.~\ref{fig:ghg_baseline}.
\par
GHG targets only static reconstruction from sparse input views;
it models humans as a set of 3D Gaussian scaffolds in the observed pose space and cannot be easily extended to dynamic scenarios. 
GHG handles Gaussian color prediction by pretraining a separate texture inpainting network.
GIGA learns to operate with appearance and geometry features simultaneously, yielding results of higher quality, both qualitatively (Fig.~\ref{fig:ghg_baseline}) and quantitatively (Tab.~\ref{tab:baselines}).
\par \noindent\textbf{Cross-Dataset Validation.}
\begin{figure}[tp]
    \centering
    \includegraphics[width=\linewidth]{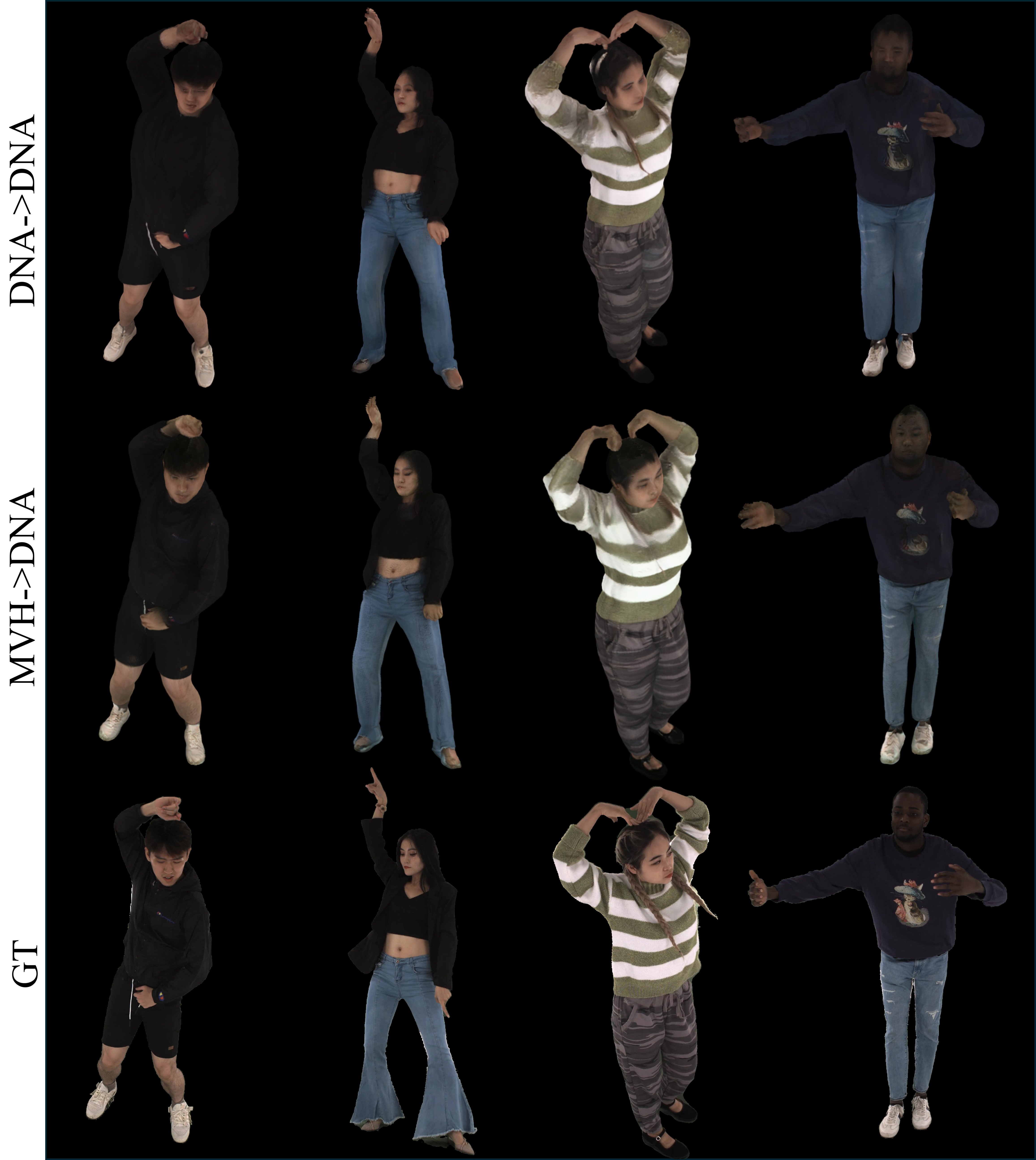}
    \caption{\textbf{Cross-Dataset Examples.} After training on MVHumanNet, GIGA successfully generalizes to novel subjects from DNA-Rendering (MVH$\to$DNA) and performs similarly to GIGA, specifically trained on DNA-Rendering (DNA$\to$DNA).}
    \label{fig:cross_dataset}
\end{figure}

To demonstrate the cross-dataset generalization ability, we train two variants of GIGA, where one is trained and tested on the DNA-Rendering~\cite{xiong_2024_cvpr} dataset, while the other variant is trained on MVHumanNet~\cite{cheng_2023_iccv} and tested on the DNA-Rendering dataset. 
As seen in Tab.~\ref{tab:cross_data_generalization} and Fig.~\ref{fig:cross_dataset}, our method is capable of generalizing across datasets with minimal quality drop, as the model trained on the MVHumanNet dataset performs comparably to the counterpart trained on the DNA-Rendering dataset, which clearly shows that our method effectively learns the prior from large datasets.
\par \noindent \textbf{Single/Sparse-View Inputs.}
\label{sparse_inputs}
To enable a single-input-view regime at test-time, for each character we precompute a static RGB texture from multiple views and concatenate it with $\mathbf{T}_{\mathbf{x}_0}$ from Eq.~\eqref{eq:method:appr} along channel dimension.
This serves as "mean" appearance information to be used for unobserved parts of the template by the network.
We retrain GIGA with this additional input and evaluate the effect of the number of views at test-time in Tab.~\ref{tab:abl_architecture}.
The 4-view setup (front, back, left, and right; \textbf{0c}) provides a full coverage of the character, whereas the 2-view setup (\textbf{0b}) only uses the front and the back views, resulting in slightly better metric values.
In the single view setup (\textbf{0a}), we only provide the frontal view, which contains information about the dynamic appearance, while the backside of the character will be reconstructed mainly based on the precomputed static texture - thus leading to reduced quantitative performance.
We report qualitative results in the supplementary material.
\subsection{Ablation Study}
\label{sec:ablations}
In this section, we discuss our design choices for the architecture and training objective of GIGA. 
We also examine the effects of different sizes of training datasets and the influence of motion tracking from sparse input views in the supplementary material.
\begin{figure}[tp]
    \centering
    \includegraphics[width=\linewidth]{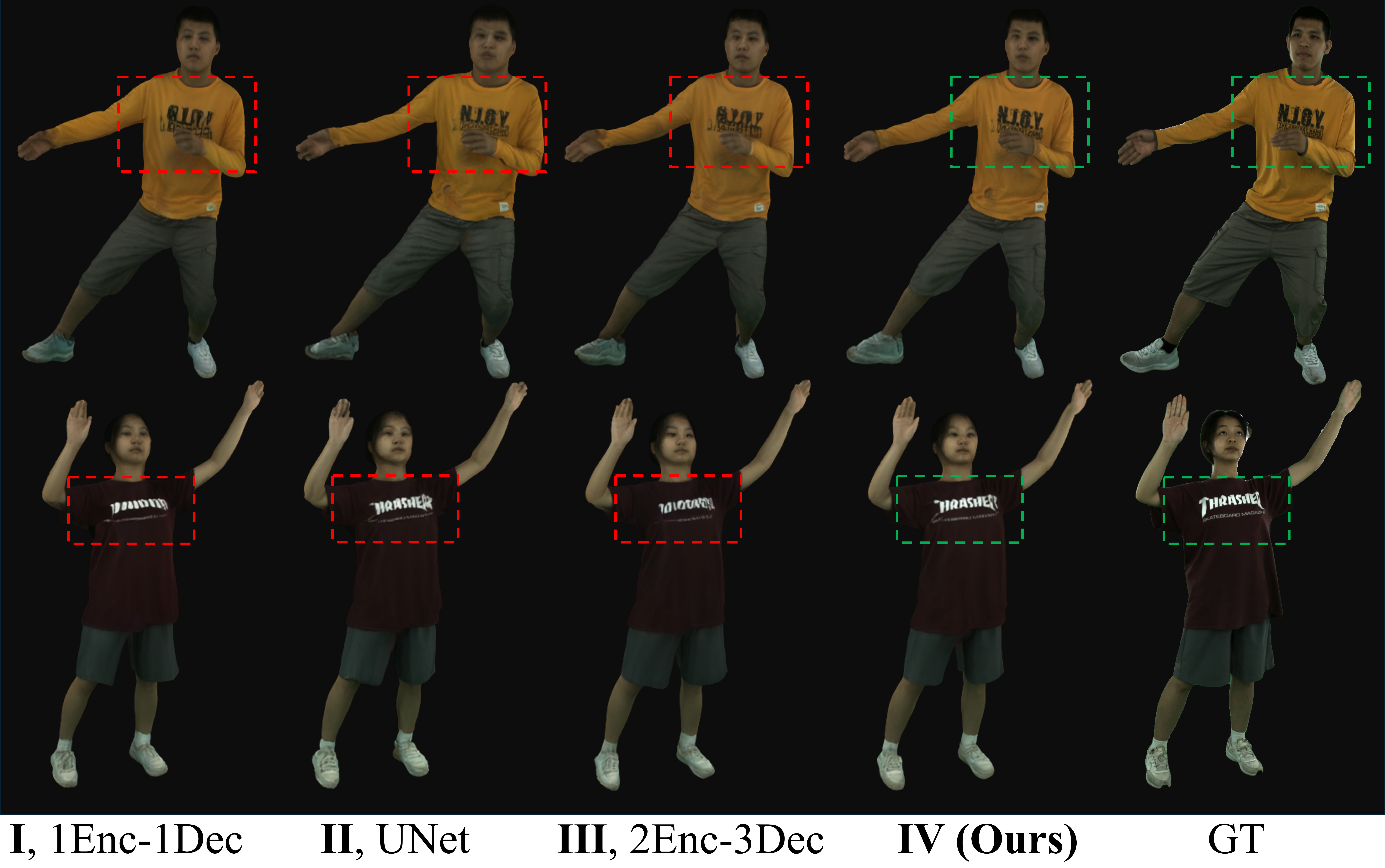}
    \caption{\textbf{Impact of Different GIGA Network Architectures.} While all proposed GIGA configurations produce reasonable results at coarse level, multiple encoding and decoding heads with additional skip-connections (\textbf{IV}, MultiHeadUNet) manages fine-grained appearance better.}
    \label{fig:arch_ablation}
\end{figure}

\par \noindent \textbf{Model Architecture.}
\label{sec:abl_model_arch}
To benchmark MultiHeadUNet (\textbf{IV}) at the core of GIGA, we propose 3 alternative architectures with approximately the same number of trainable parameters ($\simeq$90M): 
a simple encoder-decoder model (\textbf{I}), 
a conventional UNet with skip-connections between corresponding up- and downsampling blocks of the encoder and decoder (\textbf{II}), 
and a model with 2 encoders $\mathcal{E}_\mathrm{a}, \mathcal{E}_\mathrm{g}$ and 3 decoders $\mathcal{D}_\mathrm{a}, \mathcal{D}_\mathrm{p}, \mathcal{D}_\mathrm{g}$, but without skip-connections (\textbf{III}).
We observe, both quantitatively (Tab.~\ref{tab:abl_architecture}) and qualitatively (Fig.~\ref{fig:arch_ablation}), that the configuration \textbf{IV} leads to a higher quality overall, being particularly helpful at preservation of fine appearance details, observed in the input signal.
\par \noindent \textbf{Motion Conditioning.}
\label{sec:abl_model_motion}
Configurations \textbf{I}-\textbf{IV} use an MLP-based motion encoder $\mathcal{E}_\mathrm{m}$ and cross-attention motion conditioning by default. 
We additionally remove cross attention blocks from \textbf{IV}, leaving intermediate features $\mathbf{F}_\mathrm{uv}^\mathrm{a}, \mathbf{F}_\mathrm{uv}^\mathrm{g}$ without motion conditioning (\textbf{IVa}).
As it is evident from the quantitative evaluation (\ref{tab:abl_architecture}), enabling motion conditioning via cross-attention improves the visual quality of GIGA's predictions.
\par \noindent \textbf{Regularizations.}
\label{sec:abl_regularizations}
Additional geometry penalties also contribute to the quality of GIGA. 
Removing opacity-related penalties $\mathcal{L}_\alpha, 
\mathcal{L}_{\delta\mathbf{x}_0; \alpha}, \mathcal{L}_{\mathbf{s}; \alpha}$ (\textbf{IVd}), 
and then the offset penalty $\mathcal{L}_{\delta \mathbf{x}_0}$ (\textbf{IVc}) from the complete training objective~\eqref{loss:recon} leads to quality degradation (Tab.~\ref{tab:abl_architecture}), as some of the Gaussians become underconstrained. 
If no offset annealing is used during the warmup stage (\textbf{IVb}), GIGA fails to converge.
\begin{table}[t]
\centering
\footnotesize
\begin{tabular}{l|c|c|c}
\hline
Configuration & $\uparrow$ PSNR & $\uparrow$ SSIM & $\downarrow$ LPIPS \\
\hline
\textbf{0a}, 1 view & 21.32 & 0.7795 & 99.3 \\
\textbf{0b}, 2 views & \textbf{22.55} & \textbf{0.7992} & \textbf{85.4} \\
\textbf{0c}, 4 views & 22.28 & 0.7968 & 89.4 \\
\hline
\textbf{I}, Enc-Dec & 21.93 & 0.7450 & 75.4 \\
\textbf{II}, UNet & 21.94 & 0.7449 & 73.4 \\
\textbf{III}, 2Enc-3Dec & 21.88 & 0.7441 & 77.1 \\
\textbf{IV} (\textbf{GIGA}) MultuHeadUNet & \textbf{22.19} & \textbf{0.7526} & \textbf{70.2} \\
\hline
\textbf{IVa}, w/o CrossAttn & 21.69 & 0.7377 & 76.5 \\
\hline
\textbf{IVb}, w/o offset annealing & ---- & ---- & ---- \\
\textbf{IVc}, w/o offset penalty & 21.73 & 0.7454 & 71.0 \\
\textbf{IVd}, w/o opacity penalties & 21.78 & 0.7465 & 71.1 \\
\hline
\end{tabular}
\caption{
\textbf{Ablation Results for GIGA.} We examine our design choices, proposed in Sec.~\ref{sec:method}, see detailed analysis in Sec.~\ref{sec:ablations}.}
\label{tab:abl_architecture}
\end{table}

\section{Limitations}
\label{sec:limitations}
GIGA shows scalability to hundreds of multi-view videos, thanks to our efficient representation and efficient architecture, and respective generalization without sacrificing rendering quality.
Future work could explore the construction of a latent space for virtual humans to improve temporal consistency.
SMPL-X as a body template forms the basis for generalization, but it does not allow for handling non-rigid dynamics (e.g., hair and loose clothing) without additional assumptions or physics-based priors.
More advanced human shape prior that includes clothing geometry might alleviate some of these limitations~\cite{Zakharkin_2021_ICCV, grigorev_2023_cvpr, rong_2024_arxiv}.
To mitigate motion tracking issues, an end-to-end optimization of body shape and pose parameters could be a promising next step, which has already proven to be successful for face-only rendering approaches~\cite{teotia_2024_acmtog}.
\section{Conclusion}
\label{sec:conclusion}
This work presented GIGA, a generalizable, sparse image-driven 3D Gaussian human model.
Trained on a large-scale multi-view dataset, GIGA dynamically synthesizes texel-aligned 3D Gaussians from sparse input videos in a feed-forward manner.
Our approach achieves state-of-the-art generalization to unseen identities while preserving person-specific pose-dependent appearance changes thanks to our scalable architecture and efficient representation.
We believe our proposed model could benefit future research in this domain and take another step towards enabling more accessible and immersive remote collaboration.

{
    \small
    \bibliographystyle{ieeenat_fullname}
    \bibliography{main}
}
\appendix
\clearpage
\setcounter{section}{0}
\maketitlesupplementary
In this supplementary document, we first describe the process of input texture estimation in Sec.~\ref{supp:texture}.
Second, in Sec.~\ref{supp:training_details} we provide details about training of GIGA, specifying data splits (\ref{supp:training_details:data}) and the training protocol (\ref{supp:training_details:procedure}).
Third, we present training and evaluation setups for the baselines in Sec.~\ref{supp:baselines}.
Then, in Sec.~\ref{supp:input_settings} we elaborate on the test time regime of GIGA with only a monocular input.
Afterwards, we report in-the-wild examples on a monocular videos dataset in Sec.~\ref{supp:inthewild}.
Finally, we present additional ablation studies conducted for GIGA in Sec.~\ref{supp:ablations}: the effect of scaling the training dataset size and qualitative results with motion tracking from sparse views.
\par
We also report qualitative results of GIGA in the supplementary video.
\section{Projected Texture Estimation.}
\label{supp:texture}
The input views consist of rich identity-specific appearance and pose-dependent variations. 
To aggregate the identity and pose-dependent information from input views, we adopt inverse texture mapping, which projects the imagery from multiple input views to the texel space of the template mesh.

\noindent \textbf{Mesh Normalization.}
Before constructing inputs for GIGA, we normalize canonical mesh $\mathcal{V}(\boldsymbol{\theta}_0, \boldsymbol{\beta})$.
We compute the scaling factor $\rho_{\mathrm{body}} = \max \left(|\widetilde{\mathbf{x}}_0| \right)$ from canonical mesh vertices $\widetilde{\mathbf{x}}_0 \in \mathbb{R}^{N \times 3}$.
Normalized vertices $\mathbf{x}_0 = \rho_{\mathrm{body}}^{-1} \widetilde{\mathbf{x}}_0$ fit in the cubic region $\left[-1, 1 \right]^3$.
The same scaling procedure is applied to translation vectors $\pi_{k, \mathrm{o}} \in \mathbb{R}^3$ of each of $K$ cameras. 
\par 
\noindent \textbf{Partial Texture Computation.}
After articulating the normalized canonical mesh to the observed pose $\boldsymbol{\theta}_j$, we obtain posed vertex coordinates $\mathbf{x}_j$.
The next step is an initialization of the texel coordinates buffer $\mathbf{T}_{\mathbf{x}} \in \mathbb{R}^{T \times T \times 3}$.
We set $\mathbf{x}_j$ as attributes to the mesh $\mathcal{V}(\boldsymbol{\theta}_j, \boldsymbol{\beta})$ vertices 
and perform texture sampling w.r.t UV parametrization $\mathcal{M}_{\mathrm{uv}}$ to fill buffer $\mathbf{T}_{\mathbf{x}}$ with coordinates of posed texels.
In the following, we will drop the pose index $j$, assuming that all operations are performed for the observed pose.
For each input view $k$, pixel coordinates are calculated for each texel:
\begin{equation}
    \mathbf{T}_{\mathbf{x}, k}' = \Proj_k \left(\mathbf{T}_{\mathbf{x}}\right),
    \label{supp:eq:projection}
\end{equation}
where $\Proj_k$ denotes OpenGL-style projection to clip-space of view $\pi_k$.
\par 
The partial texture $\mathbf{T}_{\mathrm{uv}, k}$ is bilinearly sampled from the image $I_k$ using pixel coordinates $\mathbf{X}_{k}'$:
\begin{equation}
    \mathbf{T}_{\mathrm{uv}, k} = \GridSample \left(I_k, \mathbf{T}_{\mathbf{x}, k}'\right).
    \label{supp:eq:partial_texture}
\end{equation}
\par \noindent \textbf{Visibility Check and Texture Aggregation.}
Not every texel is observed from the view $\pi_k$. Hence, we need to discard invisible texels from the partial texture.
We first render a depth image of the body template $\mathcal{V}\left(\boldsymbol{\theta}, \boldsymbol{\beta}\right)$  and retrieve the vertex visibility buffer provided by the differentiable rasterizer~\cite{laine_2020_tog}.
After barycentric interpolating the visibility buffer $\mathcal{M}_{\mathrm{uv}}$, we obtain the visibility mask $\mathbf{V}_{\mathrm{uv}, k} \in \mathbb{R}^{T \times T}$. 
Then, we compute angle visibility scores $
\mathbf{V}^{a}_{\mathrm{uv}, k} \in \mathbb{R}^{T \times T}$:
\begin{equation}
    \mathbf{V}^{a}_{\mathrm{uv}, k} = \left( \mathbf{N} \cdot \unit\left(\pi_{k, \mathrm{o}} - \mathbf{T}_{\mathbf{x}} \right) \right),
    \label{supp:eq:angle_visibility}
\end{equation}
where $\mathbf{N} \in \mathbb{R}^{T \times T \times 3}$ are per-texel normals obtained through barycentric interpolation of the vertex normals.
$\unit$ denotes L2-normalization for per-texel viewing direction, 
and $\left( \odot \right)$ denotes dot product between vectors.
Next, we calculate indices $\overline{k}$ of partial texture with highest visibility scores
\begin{equation}
    \overline{k} = \underset{k}{\argsort} \left(\mathbf{V}^{a}_{\mathrm{uv}, k}\right).
    \label{supp:eq:visibility_indices}
\end{equation}
Finally, the (body-) pose-dependent RGB texture map $\mathbf{T}_{\mathrm{uv}}$ is computed as follows:
\begin{equation}
    \mathbf{T}_{\mathrm{uv}} = \vecgather \left( 
    \mathbf{T}_{\mathrm{uv}, k} \odot \mathbf{V}_{\mathrm{uv}, k}, \overline{k}\right)
    \label{supp:eq:texture_aggregation}
\end{equation}
where $\odot$ stands for Hadamard product and $\vecgather$ performs aggregation of individual texels specified by indices $\overline{k}$.
\section{Training Details.}
\label{supp:training_details}
\begin{table}[t]
\centering
\vspace{-2.8mm}
\begin{tabular}{l|c|c|c|c}
& \multicolumn{4}{c}{Number of training characters} \\
\hline
Metric & 100 & 250 & 500 & 970 \\
\hline
$\uparrow$ PSNR & 21.02 & 21.37 & 21.18 & 22.19 \\
\hline
$\uparrow$ SSIM & 0.7165 & 0.7291 & 0.7321 & 0.7526 \\
\hline
$\downarrow$ LPIPS & 86.9 & 82.7 & 80.1 & 70.2 \\
\hline
\end{tabular}
\vspace{-8px}
\caption{\textbf{Effect of the Dataset Size.} Increasing number of subjects in the training dataset leads to a consistent improvement in terms of generalization to novel identities.}
\label{tab:abl_dataset_size}
\vspace{-8pt}
\end{table}

\begin{figure*}[tp]
    \centering
    \includegraphics[width=1\linewidth]{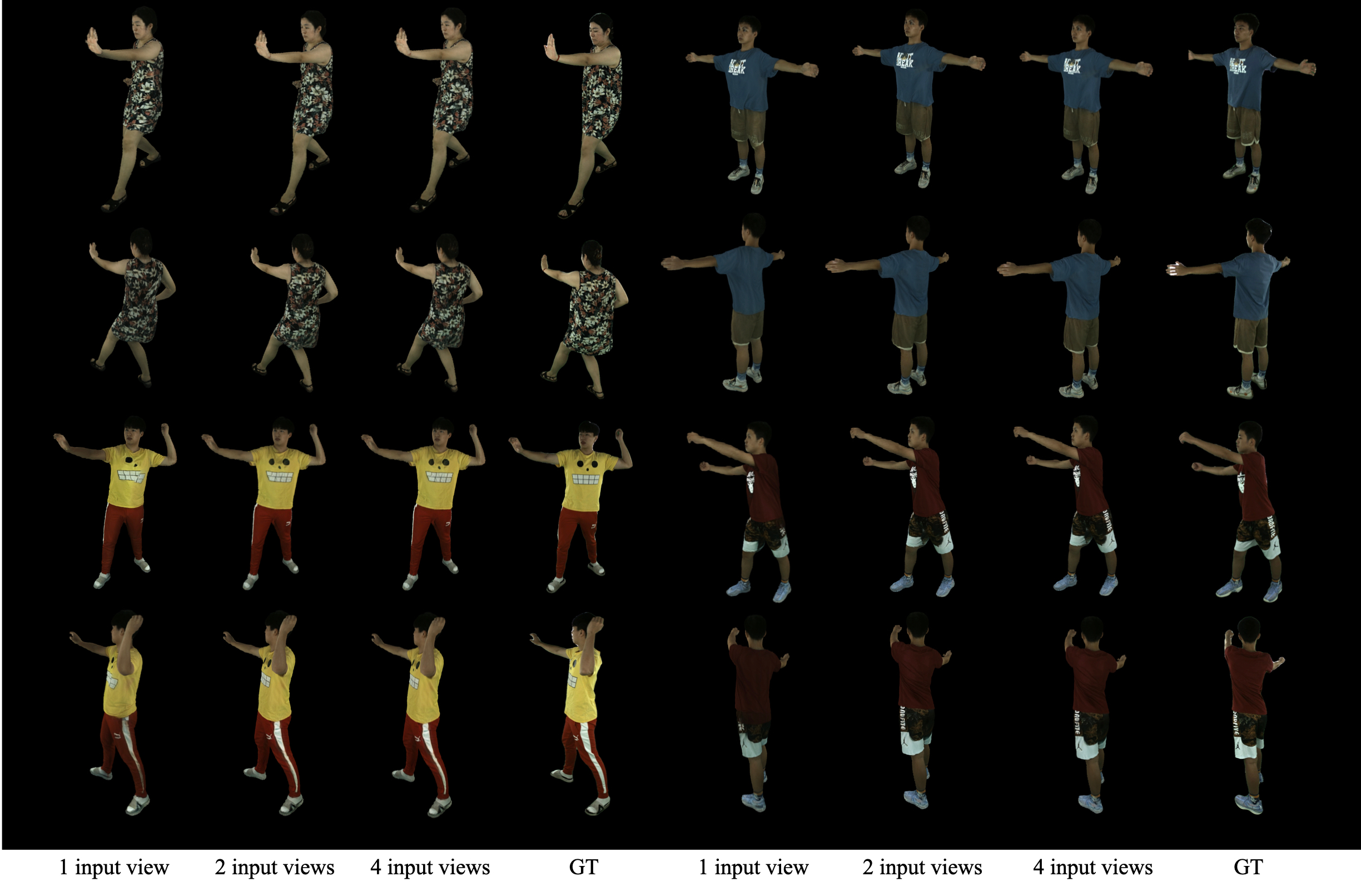}
    \caption{\textbf{Qualitative Results in Single/Sparse-View Setups.} GIGA can synthesize virtual humans from a single or a few (2-4) sparse views.}
    \label{fig:sparse_views_suppl}
    \vspace{-8pt}
\end{figure*}

\subsection{Data Splits.}
\label{supp:training_details:data}
\par \noindent \textbf{MVHumanNet.} We split MVHumanNet as follows: IDs $100000-100999$ are used for training and IDs $101000-102386$ are held out for validation and testing.
From each subject sequence, we sample every $20$th pose for training, but not more than $15$ poses per character.
From the testing part, we chose $40$ subjects for quantitative evaluation.
\par \noindent \textbf{DNA-Rendering.} For experiments with DNA-Rendering, we use the released Part-2 for training (400 subjects) and Part-1 for validation (39 subjects).
\par \noindent \textbf{THuman2.1.} Since the original release of the THuman2.0 dataset with 526 3D scans, it has been extended with additional 1919 scans and rebranded as THuman2.1.
We use the new 1919 scans for training of GIGA and corresponding baselines (GPS-Gaussian~\cite{zheng2024gps} and GHG~\cite{kwon_2024_eccv}), and use 100 old scans for validation as specified in the GHG code release.
We render multi-view images from the scans and use SMPL-X parameters provided by the authors of the dataset.
\subsection{Training Procedure.}
\label{supp:training_details:procedure}
We render $4$ random views for each input subject to compute the training objective. 
When computing perceptual loss, we sample 16 patches of varying sizes: $128 \times 128$, $256 \times 256$, and $512 \times 512$; then resize all patches to $256 \times 256$ before computing the loss, following \citet{cao_2022_tog}, to ensure scale-invariance of perceptual features.
\begin{figure*}[tp]
    \centering
    \includegraphics[width=1\linewidth]{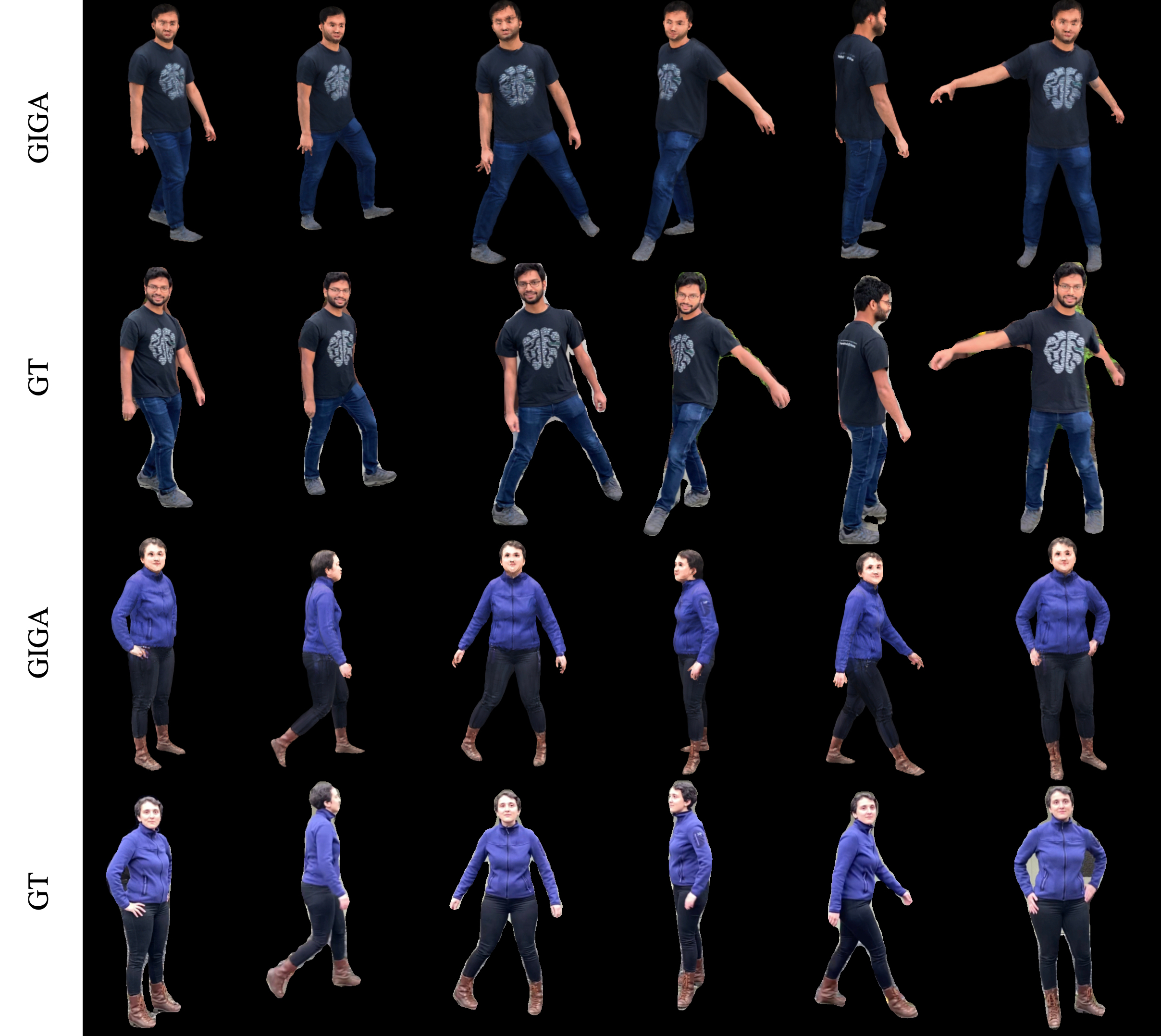}
    \caption{\textbf{In-the-Wild Examples.} Qualitative Results on the Neuman dataset.}
    \label{fig:neuman}
    \vspace{-8pt}
\end{figure*}

\section{Baselines.}
\label{supp:baselines}
\textbf{Neural Human Performer and TransHuman.}
We use the open-source implementations of both NHP~\cite{kwon_2021_neurips} and TransHuman~\cite{pan_2023_iccv} for training.
For the sake of efficiency, we replace original attention layers in baselines with their more efficient analogs~\cite{xFormers2022}.
TransHuman requires clusterization of body template vertices. 
Therefore, we follow the original codebase and cluster SMPL-X vertices using K-Means with $K=300$ clusters.
We train NHP and TransHuman using the same training/validation split of MVHumanNet as described in subsection~\ref{supp:training_details:data}.
\par \noindent \textbf{GPS-Gaussian.} 
The publicly released checkpoint of GPS-Gaussian was obtained after training in a 16-view setup.
For a fair comparison, we retrain GPS-Gaussian using 6-view setup (60 degrees angular distance between views) - we observed that GPS-Gaussian had failed to converge under sparser view setups.
At test-time, our version of GPS-Gaussian works with 2 input views, having at most 67 degrees angular difference, which amounts to approximately 5 views placed around the subject in total.
\par \noindent \textbf{GHG.}
GHG contains two models: an inpainting model for incomplete RGB textures and a 3D Gaussian regression model.
We use the checkpoint of the inpainting model provided by the authors and then finetune it jointly with the Gaussian regression model.
The latter is initialized from the released checkpoint and trained further on a larger collection of 3D scans, as specified in subsection~\ref{supp:training_details:data}.
We follow the training setup of GHG, randomly sampling 3 input views during training.
For evaluation, we select 4 views with full coverage of the subject.
\section{Single/Sparse Input Views.}
\label{supp:input_settings}
\begin{figure*}
    \centering
    \includegraphics[width=\linewidth]{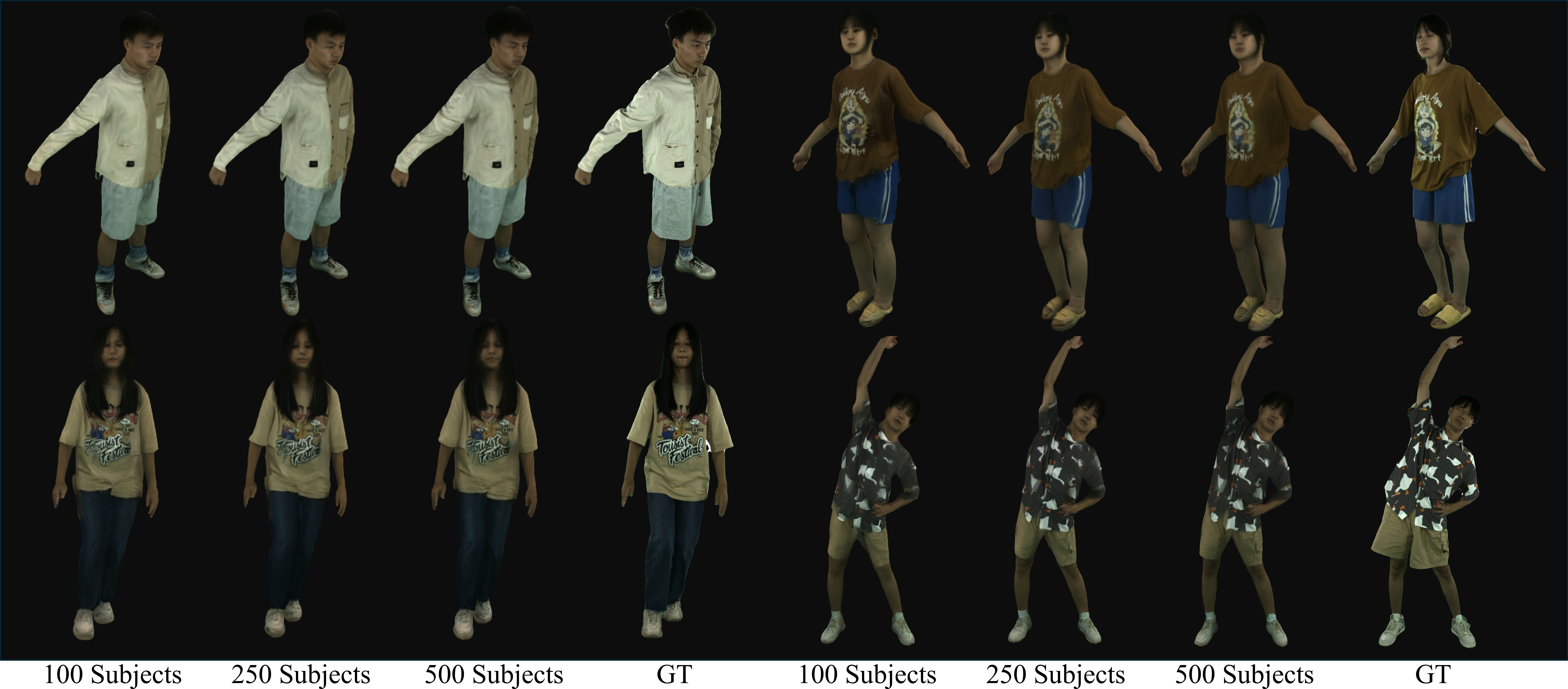}
    \caption{\textbf{Ablation on Number of Subjects for Training.} We provide qualitative results of GIGA, trained on smaller subsets of MVHumanNet~\cite{xiong_2024_cvpr}}
    \label{fig:num_subject}
\end{figure*}

As discussed in subsection~\ref{sparse_inputs}, GIGA can produce results even with a single input video.
We report qualitative results in Fig.~\ref{fig:sparse_views_suppl}, where we pick the frontal camera view for the "1 input view" setting,
frontal and back cameras for the "2 input views"; 
and frontal, back, left, and right cameras with 90 90-degree angular difference for the "4 input views" setting. 
However, since GIGA is not trained as a generative model and, hence, lacks generative capabilities, it requires some additional signal to predict meaningful 3D Gaussian parameters in the unobserved body parts.
\par
As it was briefly mentioned in subsection~\ref{sparse_inputs}, we perform a brief character calibration procedure before inference: if only a single monocular video is available, we select those frames where the various sides of the actor are visible;
if multi-view videos are available, we select 4 to 8 views with maximum coverage.
From the selected frames we compute an average RGB texture map $\overline{\mathbf{T}_{\mathbf{x}_0}}$ and combine it with the observed texture map $\mathbf{T}_{\mathbf{x}_0}$ from Eq.~\ref{eq:method:appr} to pass to the appearance encoder $\mathcal{E}_{\mathrm{a}}$.
This average texture $\overline{\mathbf{T}_{\mathbf{x}_0}}$ serves as an appearance proxy in a single/two-view input setting.
\par
Effects of this average texture are clearly visible in Fig.~\ref{fig:sparse_views_suppl}: pose-dependent clothing folds and wrinkles on the back of actors are only present in 2- and 4-view settings,
whereas in the monocular setting, appearance on the back lacks pose-dependent effects, as they were not observed in the dynamic input stream.
\section{In-the-Wild Examples.}
\label{supp:inthewild}
Neuman dataset~\cite{jiang2022neuman} contains short monocular videos of a few subjects.
We use Neural Localizer Fields (NLF)~\cite{sarandi2024nlf} to estimate SMPL-X parameters for actors in videos.
We also compute the average texture for each actor, as described in Sec.~\ref{supp:input_settings}.
Fig.~\ref{fig:neuman} presents qualitative results produced by GIGA in the monocular input setting.
Actors from Neuman are novel to GIGA, and GIGA's training dataset did not contain any monocular videos.
Despite that, renderings produced by GIGA are plausible and faithfully reconstruct observed actors. 
\section{Additional Ablations.}
\label{supp:ablations}
\par \noindent\textbf{Scaling Dataset Size.}
\label{supp:ablations:data_size}   
GIGA also benefits from training on large data collections, as reported in Tab.~\ref{tab:abl_dataset_size} and Fig.~\ref{fig:num_subject}.
While scaling a digital human rendering model to hundreds of identities presents a challenge, it leads to an improved generalization overall.
\par \noindent \textbf{Tracking from Sparse Views.}
\label{supp:ablations:num_input_views}
\begin{figure}[tp]
    \centering
    \includegraphics[width=\linewidth]{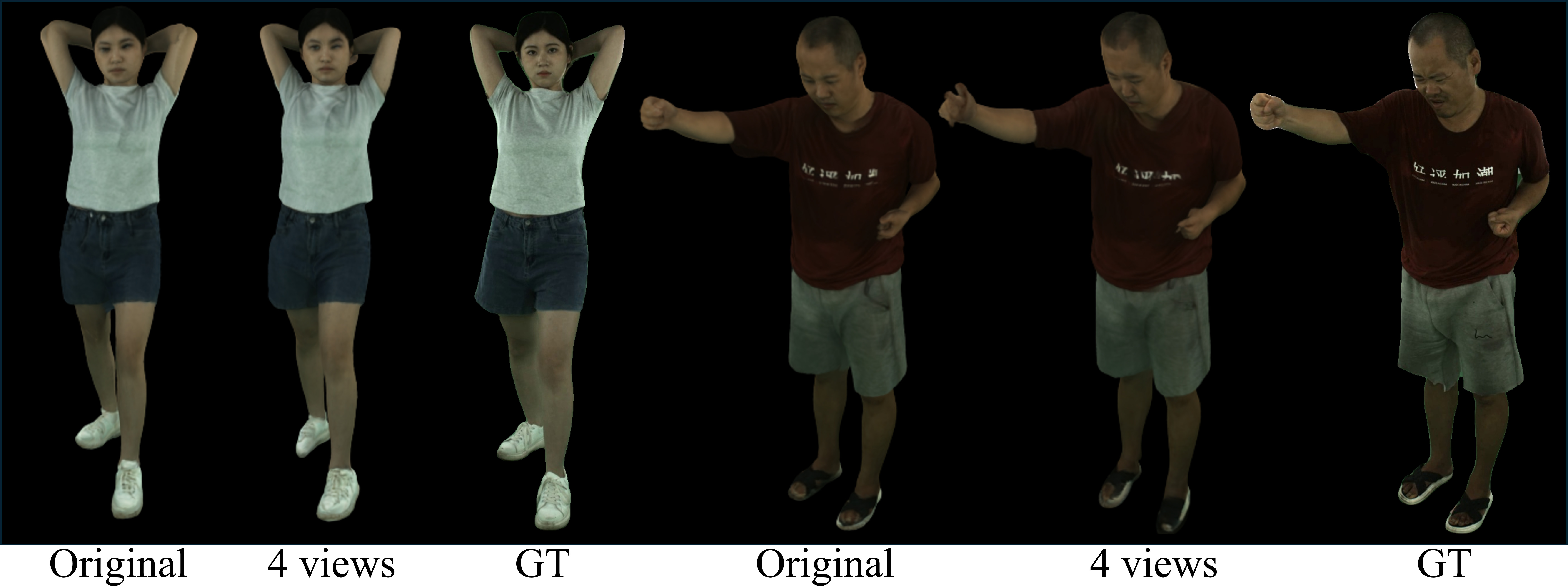}
    \caption{\textbf{Ablation on Sparsity of Tracking Cameras.} SMPL-X in MVHumanNet is estimated using tracking from 48 cameras. Retracking motion sequences from only 4 views can still yield reasonable SMPL-X templates for GIGA.}
    \label{fig:sparse_tracking}
\end{figure}

Original skeleton tracking and SMPL-X parameters in MVHumanNet are obtained from a dense set of views.
To mimic a real-world scenario, we re-track a small number of sequences from a set of 4 views and fit SMPL-X models to estimated 3D keypoints.
Qualitative results of GIGA with sparsely tracked SMPL-X template are shown in Fig.~\ref{fig:sparse_tracking}.
If sparse tracking is mostly correctly aligned with the actual human actor, then GIGA can produce results, similar to dense view tracking case.
For real-world use cases, a suitable single/sparse view motion tracking solution should be applied, for example NLF~\cite{sarandi2024nlf}, as we did for Neuman dataset in Sec.~\ref{supp:inthewild}.

\end{document}